\documentstyle[12pt]{article}

\def\fo{\hbox{{1}\kern-.25em\hbox{l}}}
\def\fnote#1#2{\begingroup\def\thefootnote{#1}\footnote{#2}\addtocounter
{footnote}{-1}\endgroup}

\renewcommand{\thefootnote}{\fnsymbol{footnote}}

\def\beq{\begin{equation}}
\def\eeq{\end{equation}}
\def\eq{\end{equation}}

\begin{document}

\newcommand{\PQ}{Peccei-Quinn~}
\newcommand{\tbqcd}{\bar{\theta}_{QCD}}
\newcommand{\lag}{{\cal L}}
\newcommand{\tr}{{\rm Tr}}
\newcommand{\fp}{f_{\pi}}
\newcommand{\fe}{f_{\eta^{\prime}}}
\newcommand{\dmu}{\partial^{\mu}}
\newcommand{\dmd}{\partial_{\mu}}
\newcommand{\epp}{\eta^{\prime}_{\perp}}
\newcommand{\ep}{\eta^{\prime}}
\newcommand{\suc}{$SU(3)_C$~}
\newcommand{\CO}{{\cal O}}
\newcommand{\CL}{{\cal L}}
\newcommand{\sulr}{$SU(2)_L \times SU(2)_R$~}
\newcommand{\ulr}{$U(2)_L \times U(2)_R$~}
\newcommand{\Dslash}{ \not \! \partial}
\newcommand{\dhg}{d_{\rm Hg}}

\begin{titlepage}

\begin{flushright}
SLAC-PUB-95-7010 \\
%CERN-TH/95-XX \\
Stanford ITP 95-22\\
hep-ph/9510220
\end{flushright}

%\vspace{.3in}
%\begin{center}
%ROUGH DRAFT, DO NOT DISTRIBUTE
%\end{center}

\LARGE
\vspace{0.05in}
\begin{center}
 Dynamical Relaxation of the \\
 Supersymmetric CP Violating Phases
\vspace{0.2in}

\large

Savas Dimopoulos\\
\vspace{.1in}
\normalsize
{\it Theoretical Physics Division,
CERN\\
CH-1211 Geneva 23, Switzerland}
\vspace{.05in}\\
{\it and} \\
\vspace{.05in}
{\it Physics Department\\
Stanford University\\
Stanford, CA  94305}\\

\vspace{.25in}

\large

Scott Thomas\fnote{\dagger}{Work supported by the Department of Energy
under contract DE-AC03-76SF00515.}\\
\vspace{.1in}
\normalsize
{\it Stanford Linear Accelerator Center\\
Stanford University\\
Stanford, CA 94309\\}

\end{center}

%\vspace{0.2in}

\vspace{0.25in}

\normalsize

A supersymmetric generalization of the Peccei-Quinn mechanism is
proposed in which two U(1) CP violating phases of the supersymmetric
standard model are promoted to dynamical variables.
This amounts to postulating the existence of spontaneously broken
global symmetries in the supersymmetry breaking sector.
The vacuum can then relax near a CP conserving point.
As a consequence the strong CP and supersymmetric CP problems
may be solved by similar mechanisms.

%\vfill
%\center{\it Submitted to Nuclear Physics B}
%\vfill

\end{titlepage}

\section{Introduction}

\baselineskip=18pt

The supersymmetric CP problem has emerged as one of the
naturalness problems for the minimal supersymmetric standard
model (MSSM).
The present bounds on the electric dipole moments of
atoms \cite{atom}, molecules \cite{molecule}, and the
neutron \cite{neutron}, restrict the the CP violating phases
discussed below to be less than ${\cal O}(10^{-2}-10^{-3})$ over
much of the MSSM parameter space \cite{bounds1,bounds2}.
A number of possible ``solutions'' to this problem have
been suggested.
In supergravity mediated supersymmetry breaking scenarios,
if the MSSM remains the effective theory up to the Planck scale,
it might be that the SUSY CP phases vanish at the Planck
scale.
The phases induced from the Kobayashi-Maskawa phase under
renormalization group evolution are then sufficiently small
at low energy \cite{smallkm}.
However, in GUT theories much larger phases are induced from
running between the Planck and GUT scales since the left and
right handed quarks and leptons generally fall within the
same multiplets \cite{GUTphases}.
Alternatively, if the superpartner masses are ${\cal O}$(TeV) the
bound on the phases is largely eliminated \cite{heavy}.
Finally, in scenarios with low energy gauge mediated SUSY breaking,
the SUSY phases simply do not arise in many classes of
models \cite{visible}.

Here we point out that promoting the phases to dynamical
variables can naturally lead to relaxation of the
vacuum to (or near) a CP conserving point.
This is an extension of the recent suggestion that the
SUSY-flavor problem may be reduced by promoting flavor rotations
among squarks to dynamical variables \cite{relaxflavor}.
In this context it can be seen as a direct extension of
the Peccei-Quinn mechanism for relaxation of the QCD vacuum
angle \cite{PQsym,kim,DFSZ}.
In fact, DFSZ type axion models \cite{DFSZ} can naturally
arise if there is a spontaneously broken global $U(1)$ symmetry
in the SUSY breaking sector.
%The $R$ symmetry in known models of dynamical supersymmetry
%breaking can in principle
%play such a role, in which case the $R$ axion
%is identified as the axion.
The existence of additional global $U(1)$ symmetries
in the SUSY breaking sector can naturally
lead to the relaxation mechanism proposed here for the SUSY CP phases.

\section{CP Violating Phases in the MSSM}

Before discussing the relaxation mechanism, let us first
review the origin of the CP violating phases in the
MSSM.
Since we are mainly interested in the flavor conserving phases
we will work under the universality anzatz.
% and discuss the effect
%of including flavor violating phases below.
With universality, four additional phases appear beyond the
Kobayashi-Maskawa phase and QCD vacuum angle of the standard model.
The first arises in the superpotential
Higgs Dirac mass parameter $\mu$,
\beq
W = \mu H_u H_d
\label{muterm}
\eeq
The remaining phases arise in the coefficients of the
soft SUSY breaking parameters $m_{\lambda}$, $A$, and $m_{12}^2$,
\beq
{\cal L} =
- {1 \over 2} m_{\lambda} \lambda \lambda
- A \left(  h_u Q H_u \bar{u}
     - h_d Q H_d \bar{d}
     - h_e L H_d \bar{e} \right)
- m_{12}^2 H_u H_d
{}~+~ h.c.
\label{softterms}
\eeq
where $\lambda$ are the gauginos, and $h_i$ the Yukawa
couplings.
Only two of the four phases are physical CP violating phases
\cite{dgh}.
This is most readily seen by noticing that in the
absence of non-gauge interactions there
are two additional flavor conserving
global $U(1)$ symmetries in the MSSM, a
Peccei-Quinn and $R$ Peccei-Quinn
symmetry \cite{rnote}.
Selection rules for the symmetries may therefore be used
if the dimensionful parameters in the couplings given above
which break the symmetries are treated as spurions with charges
assigned to compensate those of the fields, as given
in table 1.
The selection rules limit the combinations of dimensionful
parameters that can appear in a physical %(infrared finite)
amplitude.
Treating the dimensionful parameters as insertions, these are
\beq
m_{\lambda} \mu (m_{12}^2)^*,  ~~~~~
A \mu (m_{12}^2)^*,  ~~~~~
A^* m_{\lambda}
\label{combinations}
\eq
Among these there are two linearly independent phases which
may be taken to be
Arg$(A \mu (m_{12}^2)^*)$ and
Arg$(A^* m_{\lambda})$.

\begin{table}
\begin{center}
\begin{tabular}{crr}
\hline \hline
 & & \\
  & \multicolumn{1}{c}{$U(1)_{PQ}$}
  & \multicolumn{1}{c}{$U(1)_{R-PQ}$}  \\
 & & \\
\hline  \\  & & \\
$m_{\lambda}$   &  0    & $-$2  \\ & & \\
$A$             &  0    & $-$2  \\ & & \\
$m_{12}^2$      & $-$2  &  0  \\ & & \\
$\mu$           & $-$2  &  2  \\ & & \\
$H_u$           &  1    &  0  \\ & & \\
$H_d$           &  1    &  0  \\ & & \\
$Q \bar{u}$     & $-$1  &  2  \\ & & \\
$Q \bar{d}$     & $-$1  &  2  \\ & & \\
$L \bar{e}$     & $-$1  &  2  \\ & & \\
\hline \hline
\end{tabular}
\end{center}
\caption{Peccei-Quinn and $R$ charges of spurions and fields.}
\end{table}

\section{Relaxation of the Phases}

In order to motivate the relaxation solution to the SUSY CP
problem it is instructive to review the role of
nonlinearly realized global symmetries in this context.
The non-supersymmetric standard model has, at the renormalizable
level, two accidental global symmetries, namely baryon and
lepton number.
If these symmetries are realized nonlinearly there are Goldstone
bosons which couple to the associated currents.
However, if the scale of spontaneous symmetry breaking is
large enough, the Goldstone bosons decouple and there
are no effects at low energy \cite{majoron}.
The two Higgs doublet model has, in the absence of an
$m_{12}^2 H_u H_d$ term, an additional global Peccei-Quinn
symmetry at the classical level \cite{PQsym}.
This symmetry has a quantum mechanical anomaly with respect
to QCD, so $\bar{\theta}_{QCD}$ shifts under a Peccei-Quinn transformation.
If this symmetry is realized in the Goldstone mode, i.e.
Arg$(m_{12}^2)$ is a dynamical variable, the associated
pseudo-Goldstone boson (the axion) receives a potential from
the explicit breaking due to the anomaly.
It is technically natural that this potential is an extremum at
points of enhanced symmetry.
This is because if the symmetry is realized (nonlinearly or otherwise)
in the relevant degrees of freedom, then the lowest order term in
the potential near a symmetry point is bilinear in the fields.
If such points are in fact minima, then
since $\bar{\theta}_{QCD} \rightarrow - \bar{\theta}_{QCD}$ under CP,
the axion can relax to
a CP conserving point, $\bar{\theta}_{QCD} = 0$ or $\pi$.
The explicit breaking
from the anomaly comes from a marginal operator, the topological
charge density.
Since QCD is asymptotically free, the low energy long distance
dynamics {\it can} in principle
determine the potential for the axion (as is usually implicitly
assumed).
Here low energy refers to the standard model particle content with
renormalizable interactions.
It is important to note that there can be additional
explicit breakings from high energy short distance
physics, which may disturb the alignment.
High energy refers to for example GUT or
Planck scale physics, which may contain additional degrees of
freedom which do not
conserve CP with respect to the low energy theory.
However, in order for the mechanism to work, one must assume the
Peccei-Quinn symmetry is respected by the high energy physics to
a sufficiently high order in irrelevant operators \cite{breakaxion}.

%comment symmetyr in hidden sector --  coupling to visible
%sector explicitly breaks --
%phases turn out to correspond to the CP vilating phases.
% potential from low energy dynamamics minima at (or near)
%CP conserving points.

The solution of the SUSY-CP problem we propose is to
promote the phases appearing in (\ref{muterm}) and
(\ref{softterms}) to dynamical variables.
The soft terms of course arise from couplings with the
SUSY breaking sector.
Since the $\mu$ term must be of order the weak scale, the
only reasonable assumption is that it too arises from
a coupling to the SUSY breaking sector.
Promoting the phases to dynamical variables therefore amounts
to postulating the existence of spontaneously broken
global symmetries in the SUSY breaking sector.
%Spontaneous broken global symmetries are common in
%models of dynamical SUSY breaking.
%In fact, all know models of SUSY breaking which rely
%on a dynamically generated superpotential contain
%an accidental $R$ symmetry which is spontaneously broken.
The phases in ({\ref{muterm}) and (\ref{softterms}) are then
the Goldstone bosons of these nonlinearly realized symmetries.
In general all the phases need not be dynamical.
However, for now we make the ``maximal'' assumption that
all four phases are dynamical and investigate the consequences.
As we show below, one phase is analogous to baryon and lepton
number in that it is respected at the renormalizable level by
couplings to the visible sector.
The Goldstone boson which could be associated with this symmetry
therefore has no effect at low energy if the SUSY breaking scale
is large enough.
A second symmetry is analogous to the Peccei-Quinn symmetry
in the two Higgs doublet model, and can lead to the
Peccei-Quinn mechanism for the relaxation of
$\bar{\theta}_{QCD}$ \cite{ellis}
(if respected to high enough order).
The remaining two phases are the physical SUSY CP violating phases.
As we show below
any symmetries associated with these turn out
to be explicitly broken by couplings with the visible sector.
Integrating out the visible sector then produces a potential with
minima near CP conserving points for the phases.
This leads to a relaxation mechanism analogous to the
Peccei-Quinn mechanism.

%put a note somewhere about which phases get a potential and which
%dont . the ones that correspond to the physical CP violating
%phases have a potential cos.,,  therefore relax to CP conserving
%point.

The relaxation mechanism requires that %the Goldstone bosons
%associated to the phases (\ref{combinations}) are lifted and that
the vacuum energy
be a minimum at (or near) a CP conserving point
for the phases (\ref{combinations}).
The global symmetries associated with the phases
must therefore be explicitly broken by some couplings.
The possible terms appearing in the vacuum energy
which depend on the phases are limited by the selection
rules given above.
Treating the dimensionful parameters as background spurions
and including the Higgs bosons, which acquire an expectation
value at low energy, the possible
phase dependent terms are
\beq
\begin{array}{cc}
m_{\lambda} \mu H_u H_d  &  m_{\lambda} \mu (m_{12}^2)^*  \\
   & \\
A \mu H_u H_d            &  A \mu (m_{12}^2)^*  \\
   & \\
A^* m_{\lambda} H_u^* H_u & A^* m_{\lambda} H_d^* H_d \\
   & \\
A^* m_{\lambda} \Lambda^2 &  \\
\end{array}
\label{vacterms}
\eq
where each term appears with $+~h.c.$, and
$\Lambda$ is a dimensionful scale discussed below.
In general there are contributions to the terms above
from both long and short distance physics.
The long distance contributions from light degrees of freedom come from
diagrams such as those given in Figs. 1-4.
%First consider the contribution from integrating out low energy
%degrees of freedom.
%Examples of such vacuum diagrams are given in Figs. 1-4.
Just on dimensional grounds all the terms
in (\ref{vacterms}), except the last one,
correspond to marginal operators.
For the marginal operators, at worst, all
logarithmic energy scales contribute equally to the coefficients.
The contributions from integrating out light degrees of freedom can
therefore dominate the short distance contributions by
${\cal O}(\ln(\Lambda^2/m_W^2))$, where $m_W$ is the weak scale
and $\Lambda$ is the scale
at which SUSY breaking is transmitted to the visible
sector ($\Lambda \sim M_p$ in hidden sector models).
In this case the breaking of the global symmetries can come
mainly from the %coupling of the SUSY breaking sector to the
visible sector, and need not be
particularly sensitive to short distance physics.
This relative insensitivity to short distance physics
is in contrast to the situation for dynamical squark flavor
matrices \cite{relaxflavor}, dynamical Yukawa
couplings \cite{relaxyukawa,nambu}, or a dynamical determination of
the SUSY breaking scale in no-scale type models \cite{noscale,LHC};
in these cases the potential is
quadratically sensitive to the short distance physics \cite{bpr}.
Likewise here, the last term in (\ref{vacterms}) corresponds to
a relevant operator.
Integrating out light degrees of freedom therefore gives
(in the absence of a regulator) a
quadratically divergent contribution to the operator,
proportional to $\Lambda^2$.
An example of such a three loop diagram is obtained from Fig. 3.
with $H_u$ contracted with $H_u^*$.
Because of the quadratic divergence, this contribution to the
vacuum energy is very sensitive to the short distance physics.
This sensitivity implies
the precise description of the short distance contribution is
in fact scheme dependent.
For example, in dimensional regularization there are no quadratic
divergences, and the $\Lambda^2$ piece comes from the matching
conditions at the scale $\Lambda$.
The potential for the phase Arg$(A^* m_{\lambda})$ is
therefore essentially determined by physics at the scale $\Lambda$.

Now consider the form of the potential
arising from (\ref{vacterms}) for the
Goldstone bosons associated to the CP violating phases
in (\ref{combinations}).
First note that in the ground state the $m_{12}^2 H_u H_d$
term in the potential fixes
Arg$(H_u H_d)$ = $-$Arg$(m_{12}^2)$.
Ignoring for the moment the Kobayashi-Maskawa phase and
any flavor changing phases,
the long distance contributions
of the types given in Figs. 1-4
to the marginal operators then all go like
\beq
\sum_i
c_i m_W^4 ~\cos(\phi_{\alpha} + \delta_i)
\label{sumlong}
\eq
where $c_i$ is the magnitude of the $i$th diagram,
%$\delta_i = 0$ or $\pi$,
and $\phi_{\alpha}$ = Arg$(m_{\lambda} \mu (m_{12}^2)^*)$ or
Arg$(A \mu (m_{12}^2)^*)$.
Since the only CP violating phase in the lowest order diagrams
is the phase $\phi_{\alpha}$ itself, $\delta_i = 0$ or $\pi$.
The lowest order long distance contributions to the vacuum
energy therefore
necessarily have minima at CP conserving
points.
%Some of the extrema can be minima, which is in fact likely
%as a single diagram typically gives the dominant
%long distance contribution, and therefore dominates the
%sum in (\ref{sumlong}).
%(although in general there may also be minima at other points).
%For example, diagrams involving Yukawa couplings, such
%as those in Figs. 2 and 3, are dominated by the top.
%As argued above, for the marginal operators the long distance
%can dominate the short distance contributions.
For the relevant operator the lowest order contributions go like
\beq
\sum_i
c_i m_W^2 \Lambda^2 ~\cos(\phi +\delta_i)
\eq
where $\phi$ = Arg$(A^* m_{\lambda})$.
Again, the long distance contributions, such as that
in Fig. 3, give $\delta_i = 0$ or $\pi$.

The short distance pieces, however, in general have arbitrary
$\delta$.
In order to proceed without simply assuming a tuning of the short
distance phases
we must therefore assume that the global
symmetry associated with Arg$(A^* m_{\lambda})$ is realized in
the short distance physics at the scale $\Lambda$.
In addition we must assume that the
short distance physics has
the same definition of CP as the long distance physics, so that the
potential has extrema at
the CP conserving points,
$\delta = 0$ or $\pi$ \cite{relaxnote}.
%As this is a point of enhanced symmetry it is of course technically
%natural to postulate that the potential
%is a minimum there \cite{relaxnote}.
This could occur for example if CP is a symmetry of the full theory,
and only broken spontaneously below the scale $\Lambda$.
In string theory, where CP is a symmetry \cite{cpsym},
with hidden sector SUSY breaking this could occur if the
scale of spontaneous CP violation
is between the Planck and intermediate
SUSY breaking scale $M_I \sim \sqrt{m_W M_p}$.
Note that without the relaxation mechanism the soft parameters and
$\mu$ would not in general be real in this case.
With these assumptions about the short distance physics,
since the combinations of phases that
appear in the vacuum energy are precisely those that appear
in any CP odd observable, there is a ground state in which all
physical amplitudes (proportional to the terms in (\ref{combinations}))
are CP conserving.
This can also be seen by
starting from the original basis for the phases in (\ref{muterm}) and
(\ref{softterms}).
$U(1)_{PQ}$ and $U(1)_{R-PQ}$ redefinitions may always
be used to transform to a basis in which any two of the phases vanish,
for example
$\phi_A = \phi_{m_{12}^2} =0$.
In the CP conserving ground state the alignment of
Arg$(A^* m_{\lambda})$ then forces $\phi_{m_{\lambda}}=0$ or $\pi$,
and the alignment of Arg$(A \mu (m_{12}^2)^*)$ forces
$\phi_{\mu} = 0$ or $\pi$.

The alignment of the phases described above can be disturbed
in a number of ways.
Higher order loop diagrams of light degrees of freedom
can be proportional to products of the invariants (\ref{vacterms}),
and therefore have potentials proportional
to $\cos( n \phi_{\alpha} \pm n^{\prime} \phi_{\beta} + \delta )$.
These however are suppressed by at least two additional loop factors
%${\cal O}(\alpha / 4 \pi)^2$,
and do not shift the CP conserving minima.
The Kobayashi-Maskawa phase can in principle shift the minimum
of the potential from a CP conserving point.
To form the Jarlskog invariant
%$J= {\rm Im} (V_{us}V_{cs}^*V_{cd}V_{ud}^*)$
$J= {\rm Im} (V_{ud}V_{td}^*V_{tb}V_{ub}^*)$
however requires
at least four  $SU(2)$ gauge couplings.
This requires at least two additional loops compared with the
lowest order diagrams, and is therefore down by
at least ${\cal O}((\alpha / 4 \pi)^2 J)$.
GIM suppression among the squarks would reduce this contribution
even further.
The Kobayashi-Maskawa phase therefore does not significantly shift
the minima from CP conserving points.
The alignment can also be disturbed by
explicit breaking by the short distance physics of
both CP symmetry and
the
global symmetries in the hidden
sector which make the phases dynamical.
Here, just as for the Peccei-Quinn mechanism, we must assume that
the minimum is not shifted to high enough order in irrelevant operators.
However, since the bound on the phases is
${\cal O}(10^{-2}-10^{-3})$ this is not nearly as restrictive as
for the axion.
%since the phases only need to be smaller than
%${\cal O}(10^{-2}-10^{-3})$ this is not very restrictive.

\section{The Physical (Pseudo)-Goldstone Bosons}

%Now consider the properties of the physical Goldstone
%bosons.
The physical
pseudo-Goldstone bosons are related to the phases
%on the magnitude of contributions to the potential (\ref{vacterms})
by $\phi_{\alpha} = G_{\alpha} / f_{\alpha}$.
%where
%$\phi_{\alpha}$ is one of the two independent phases in (\ref{combinations}).
The decay constants $f_{\alpha}$ are
essentially the expectation values for fields in the
SUSY breaking sector which transform under the symmetries.
All the Goldstone bosons couple to the visible sector
through dimensionful couplings with a suppression of
$1/f$,
\beq
W = |\mu| e^{i Q_{\mu \alpha} \phi_{\alpha} / f_{\alpha} } H_u H_d
\eq
$$
{\cal L} = - \frac{1}{2} |m_{\lambda}|
   e^{i Q_{\lambda \alpha} \phi_{\alpha} / f_{\alpha} }
     \lambda \lambda
- |A| e^{i Q_{A \alpha} \phi_{\alpha} / f_{\alpha} }
   \left(  h_u Q H_u \bar{u}
     - h_d Q H_d \bar{d}
     - h_e L H_d \bar{e} \right)
$$
\beq
- |m_{12}^2| e^{i Q_{ud \alpha} \phi_{\alpha} / f_{\alpha}}
H_u H_d
{}~+~ h.c.
\label{goldcouplings}
\eq
where the $Q_{i \alpha}$ depend on the global charge assignments in
the SUSY breaking sector.
For $f$ above the weak scale the Goldstone bosons are
essentially ``invisible'' to laboratory experiments.
The masses for the Goldstone bosons corresponding to SUSY CP violating
phases depend on the magnitude of (\ref{vacterms}) and the decay
constant.
The mass for the linear combination lifted by the marginal operators
is then $m_G \sim \left( \sqrt{ \alpha / 4 \pi} \right) m_W^2 / f$,
where $\alpha/ 4 \pi$ counts the loop factor, while
the mass for the combination lifted by the relevant operator is
$m_G \sim m_W \Lambda / f$.
In the universal case with four dynamical phases,
due to the
$U(1)_{PQ}$ and $U(1)_{R-PQ}$ symmetries of the $\mu$ and
soft terms, the
two linear combinations of phases orthogonal to the physical
SUSY CP violating phases do not receive a
perturbative potential from coupling
with the visible sector (\ref{vacterms}).
One linear combination is anomaly free with respect to QCD, and
so receives a potential only from any explicit breaking from short
distance physics.
The other linear combination is anomalous and receives a
potential from the QCD topological  charge density.
If this symmetry is respected by the short distance physics to
high enough order in irrelevant operators, the Goldstone boson
just acts as an invisible axion with mass
$m \sim m_{\pi} f_{\pi} / f$,
thereby solving the strong CP
problem.
The Peccei-Quinn mechanism can therefore be wedded with the
proposal to solve the SUSY CP problem by postulating global
symmetries in the SUSY breaking sector.

Depending on the mechanism which transmits SUSY breaking to the
visible sector, $f$ could be anywhere between just above the
weak scale to the Planck scale.
One possibility is a renormalizable hidden sector in which SUSY is
broken in the flat space limit, but transmitted to the visible
sector by gravitational strength interactions.
The scalar expectation values in the hidden sector are
then of order the SUSY breaking
scale, $f \sim M_I \sim \sqrt{m_W M_p}$.
If, as suggested above, both the axion and the pseudo-Goldstone bosons
responsible  for relaxing the SUSY CP phases all arise from
spontaneously broken global symmetries in the SUSY breaking sector,
then apparently this type of hidden sector naturally
gives a decay constant
in the ``axion window'' allowed by astrophysical and cosmological
bounds on the axion \cite{axionbounds}.
{}From (\ref{goldcouplings}) it is apparent that this axion couples
both as a hadronic axion \cite{kim} through the gluino mass term and
as a DFSZ axion \cite{DFSZ} through the scalar Higgs term.
Note that the small coupling introduced in the original DFSZ
models by hand appears here automatically as $m_W/f$.
With this type of hidden sector
the pseudo-Goldstone bosons associated with the SUSY CP violating
phases do not lead to excessive cooling of astrophysical systems, and
are heavier than the axion and
therefore not overproduced in the early universe.

With a renormalizable hidden sector it turns out that
$A$ terms arise only from Kahler potential couplings and
are always real \cite{smallmu}.
In this case there are only three possible dynamical phases
in the universal case, thereby eliminating the
(potentially massless) anomaly free linear combination.
One may be tempted to identify one of the required Goldstone bosons
with the $R$ axion which plays a role in all known
models of dynamical SUSY breaking based on a nonperturbative
superpotential.
However, for a renormalizable hidden sector, cancelation of
the cosmological constant by adjustment of the superpotential, explicitly
breaks the $R$ symmetry \cite{bkn}.

It is also worth noting that in renormalizable hidden sector models
the heavier pseudo-Goldstone boson
has a sizeable mass, $m \sim M_I$.
This is because of the sensitivity of the relevant operator
to physics at the scale $\Lambda \sim M_p$.
The mass in other types of SUSY breaking sectors is parametrically
less than the intrinsic SUSY breaking scale.

Another possibility is a nonrenormalizable hidden sector in which
SUSY is restored in the flat space limit.
In this case the fields can have $f \sim M_p$.
With a nonrenormalizable hidden sector
it is possible to cancel the cosmological constant by adjusting the
Kahler potential without explicitly violating $R$ symmetry.
It it is then possible in principle to identify
the $R$ axion with one of the Goldstone bosons.
If both the axion and pseudo-Goldstone bosons arise from
spontaneously broken global symmetries in this
type of hidden sector, they will generally be overproduced
in the early universe.
They may be diluted by a period of very late inflation
however \cite{dimhall}.

One interesting consequence of the relaxation mechanism for
the SUSY CP phases %$f \sim M_p$
is the possibility
of a long range force. %from the pseudo-Goldstone
%boson which receives a mass from the relevant operators.
%$m \sim m_W^2 / M_p$.
If the minimum of the potential does not align precisely with a
CP conserving point, $\phi_0 \neq 0$ or $\pi$,
the interactions (\ref{goldcouplings}) lead to scalar couplings
of the pseudo-Goldstone bosons
to matter proportional to mass.
This gives rise to a coherent potential between macroscopic
bodies
\beq
V \simeq - g^2 \phi_0^2 \frac{m_i m_j}{4 \pi f^2} \frac{e^{-mr}}{r}
\eq
where
$g \sim {\cal O}(1)$ is a dimensionless coupling which depends on the
charges appearing in (\ref{goldcouplings}),
$\phi_0$ is the minimum of the phase potential mod $\pi$, and
$m^{-1}$ is the Compton wavelength.
The (lighter) Goldstone boson, which receives a mass from the marginal
operators, gives the longest range force.
For $f \sim M_p$ the Compton wavelength is ${\cal O}(10^{-1} - 100)$ cm,
and is weaker than gravity by roughly the factor $\phi_0^2$.
As argued above, the shift in the potential from long distance physics
can be quite small in the universal case,
but short distance physics or non-universality can
in principle disturb the alignment.
Notice that for $\phi_0 \neq 0$
the magnitude of CP odd observables, such as
electric dipole moments, is correlated with the strength of the long
range force.
This may be the best laboratory signal for the relaxation mechanism.

%For a renormalizable hidden sector with $f \sim M_I \sim \sqrt{m_W M_p}$
%this gives masses of order $m_W^2 / M_I$ and $M_I$ respectively.

%The shift of the minimum of the lighter of these Goldstone bosons
%due to Planck suppressed operators which explicitly break the
%associated global symmetry {\it and} CP symmetry is
%$\delta \phi \sim f^{n}/(M_p^{n-4} m_W^4)$ where $n$ is the dimension
%of the operator.
%For a renormalizable hidden sector, requiring that this not
%destroy the alignment
%implies $n > 8$.

%Although the two SUSY CP phases obtain potentials from
%the terms (\ref{vacterms}) the two remaining phases
%in (\ref{muterm}) and (\ref{softterms}) do not.
%If these phases

%DFSZ....
%small coupling to $H_u H_d$ introduced by hand, is
%just a result of the hierarchy between the weak and
%hidden sector SUSY breaking scales. a result

\section{Conclusions}

Promoting the SUSY CP phases to dynamical variables allows
the possibility that the vacuum can relax to a ground
state at or near
a CP conserving point for these phases.
The long distance contribution to the potential for
the flavor conserving phases can in fact have a minimum near
a CP conserving point.
Although not calculable in the low energy theory, it is
technically natural that the short distance
potential also lies near a CP conserving point.
Promoting the phases to dynamical variables amounts to postulating
the existence of spontaneously broken global symmetries in the
SUSY breaking sector.
This requires protecting (in the limit of decoupling the
visible sector) some compact flat directions
in the SUSY breaking sector,
which is of course possible with symmetries.
This is in contrast to some other mechanisms for dynamically determining
low energy parameters which require protecting noncompact flat
directions in the presence of SUSY breaking.
Noncompact symmetries can do this at the classical level, but are
generally violated quantum mechanically.
Also note that since the relaxation of these phases requires
global symmetries in the SUSY breaking sector, other CP violating
phases unrelated to this sector, such as the Kobayashi-Maskawa phase,
of course need not be dynamical.
Finally,
the mechanism for the relaxation of the SUSY CP violating phases
is a generalization of
the Peccei-Quinn mechanism for the solution of the strong
CP problem.

After this work was completed
it was brought to our attention that
Ref. \cite{choi} considered the possibility of dynamical
phases in models of
moduli dominated supersymmetry breaking.

\newpage

\large
\noindent
{\bf Figure Captions}
\vskip.5in
\normalsize

\begin{description}

\item[Figure 1.] One loop contribution to the vacuum energy
proportional to $m_{\lambda} \mu H_u H_d + h.c.$
Contracting $H_u$ and $H_d$ gives a two loop contribution
proportional to $m_{\lambda} \mu (m_{12}^2)^* + h.c.$

\vskip.15in

\item[Figure 2.] One loop contribution to the vacuum energy
proportional to
$A \mu H_u H_d + h.c.$
Contracting $H_u$ and $H_d$ gives a two loop contribution proportional
to $A \mu (m_{12}^2)^* + h.c.$

\vskip.15in

\item[Figure 3.] Two loop contribution to the vacuum energy
proportional to
$m_{\lambda} \mu H_u H_d + h.c.$ and
$A^* m_{\lambda} H_u^*H_u + h.c.$
Contracting $H_u$ and $H_u^*$ gives a quadratically divergent
contribution proportional to $A^* m_{\lambda} + h.c.$.

\vskip.15in

\item[Figure 4.] A three loop contribution to the
vacuum energy proportional to $A \mu (m_{12}^2)^* + h.c.$

\end{description}


\begin{thebibliography}{20}
\bibitem{rnote} Under an $R$ transformation
the scalar, fermionic, and auxillary components of
a chiral superfield transform as
$\phi \rightarrow e^{i \alpha R} \phi$,
$\psi \rightarrow e^{i \alpha (R-1)} \psi$, and
$F    \rightarrow e^{i \alpha (R-2)} F$ respectively, where
$R$ is the $R$ charge.
The superpotential has $R$ charge $R(W)=2$, and
a gauge superfield has $R$ charge
$R(W^{\alpha})=1$.
\bibitem{relaxnote} It is worth noting that the
technically natural option of postulating a minimum of the
short distance potential
at an enhanced symmetry point can also be applied to the
case of dynamical squark flavor rotations, but does not appear
to be available for dynamical Yukawa couplings or
the dynamical determination of the
SUSY breaking scale in no-scale type models.
\end{thebibliography}
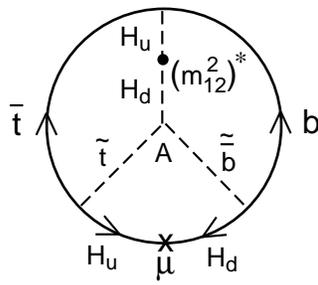
\end{document}